\documentclass{article}

\usepackage{mathptmx}
\usepackage{amsmath}
\usepackage{graphicx}
\usepackage[round]{natbib} %For better citing.
\setcitestyle{aysep={}} %to remove comma from between author and year

\usepackage{epigraph}
\begin{document}
	
	\title{Spontaneous Collapse of the Wavefunction: \\ a Testable Model Motivated by Discrete Physics}
	\author{Martin J. Leckey and Adrian P. Flitney
	}
	
	\date{\today}
	
	% \pagenumpos{\pnbc}
	\maketitle
	% 	wln
	
	%Commands
	\newcommand{\x}{{\bf x}}
	\newcommand{\y}{{\bf y}}

\begin{abstract}
	A modified form of quantum mechanics which includes a new mechanism for wavefunction collapse is proposed.
	The collapse provides a solution to the quantum measurement problem.
	This modified quantum mechanics is shown to arise naturally from a fully discrete physics,
	where all physical quantities are discrete rather than continuous.
	We compare the theory to the spontaneous collapse theories of Ghirardi, Rimini, Weber and Pearle, and argue that the new theory lends itself well to a realist interpretation of the wavefunction. 
	%\keywords{complexity \and discrete physics \and measurement problem \and non-linear dynamics \and quantum mechanics \and wavefunction collapse}
\end{abstract}

\section{Introduction}
\label{Sec:Intro}
\epigraph{
	Glory is like a circle in the water,\\
	Which never ceaseth to enlarge itself\\
	Till by broad spreading it disperse to nought.}
{Shakespeare, {\em Henry VI, part 1}}

We propose a modified quantum mechanics
that includes an outline of a solution to the `measurement problem.'
%However, the modifications that are introduced are not motivated merely by a desire to solve the measurement problem.
The theory is shown to arise naturally from a fully discrete physics, where all physical quantities,
including the magnitude of the wavefunction, are discrete.
Since normal wavefunction evolution leads to an expansion or spreading with time,
eventually the wavefunction will grow to such an extent that the average magnitude of the wavefunction
threatens to fall below the minimum magnitude representable in the discrete physics under consideration,
thus threatening to disappear altogether.
Our key idea is that, since the wavefunction does not vanish, 
the existence of a minimum magnitude provides a reason, independent of any consideration of measurement,
to suppose that the wavefunction will `collapse',
suddenly reducing in size, when it reaches a critical volume in configuration space.
%How this form of collapse leads to a solution to the measurement problem will be explained in what follows.

In nature, there would appear to be no truly isolated systems,
that is, systems with no interaction at all with their environment.
According to standard quantum mechanics,
this means that the universe is most accurately represented by a single wavefunction.
In our model,
which has wavefunctions of finite volume,
the number of particles in the wavefunction is related to the strength of interaction between the particles.
%In our model, the number of particles per wavefunction is fewer when the particles interact more weakly;
%wavefunctions could have as little as a single particle.
%These particles will be able to spread widely before the critical volume is reached.
%When the particles interact strongly there are more particles per wavefunction,
%and the volume of the wavefunction in configuration space rises exponentially with the %number of particles.
%In this case, the critical volume will be reached
%^+ while the particles still remain relatively localised in space
%This means this volume will reach the critical level when its particles are still
%relatively localised in space.
%This explains why particles in macroscopic systems remain localised..
%In the model discussed here, not only do wavefunctions spontaneously %collapse,
%but there are also many separate wavefunctions in the universe.
%When particles interact strongly there are more particles per wavefunction,
%so the volume in configuration space rises increasing the probability of a %collapse.
%This explains why macroscopic systems remain localised.
%Weakly interacting systems will have as little as a single particle,
%occupying a small volume of configuration space
%and hence will collapse less frequently.
%some representing as little as a single particle.
%Wavefunctions representing a large number of strongly interacting particles
%will reach the critical volume in configuration space more quickly,
Where there are a large number of strongly interacting particles,
they would be represented by a single wavefunction
that will quickly reach the critical volume and collapse,
leading to localization of macroscopic objects,
while wavefunctions representing fewer particles will expand more widely before collapsing.
This is the central idea of our solution to the measurement problem.

Our model permits a realist interpretation of the wavefunction
in which it is not merely a mathematical tool for predicting outcomes of experiments or observations,
but is also a physical feature of the universe in its own right.
Furthermore, since the theory makes novel empirical predictions, it is in principle testable.

The theory has some features in common with two closely related theories of wavefunction collapse that have been proposed with the intention of providing a solution to the measurement problem.
These are the theory of Ghirardi, Rimini, and Weber (1986, 1988) (GRW);
and the Continuous Spontaneous Localization (CSL) theory \citep{gpr1990}. 
In the following section, we give a brief account of the measurement problem.
An introduction to existing collapse theories is presented in Sect.\ \ref{Sec:ExistCollapse}.
In Sect.\ \ref{Sec:DisPhys} we discuss discrete physics and demonstrate how discretization leads naturally to a consideration of wavefunction collapse.
Section \ref{Sec:CCQM} introduces our modified quantum mechanics
%which we call Critical Complexity Quantum Mechanics (CCQM),
which we show can lead to a solution to the measurement problem
(Sects.\ \ref{Sec:Collapse}--\ref{Sec:SolnMeasProb}).
The dynamics of the new model are further explored in Sects.\ \ref{Sec:PropCollMech}--\ref{Sec:NonLinear}.
Comparison between the new theory and existing spontaneous collapse models
is given in Sect.\ \ref{Sec:Advantage}.
Section \ref{Sec:WfnReal} explores wavefunction realism
and Sect.\ \ref{Sec:Constraints} discusses observational constraints on the theory,
followed by concluding remarks.
The new theory gives rise to novel observational predictions.
Confirmation of these predictions could in principle empirically differentiate this collapse theory from both standard quantum theory and other collapse models.
%Further details about novel predictions of the theory are given in [removed to maintain anonymity].
%\citet[pp.\ 17--25]{leckey2023}.

\section{Measurement Problem}
\label{Sec:MeasProb}
In the orthodox (Dirac-von Neumann) formulation of quantum mechanics \citep{dirac1930,vonneumann1955},
the time-evolution of a system is described by the (deterministic) Schr\"{o}dinger equation.
However, when a `measurement' takes place the normal evolution of the state is suspended and the system changes indeterministically,
abruptly `collapsing', resulting in a determinate measurement outcome.
This gives rise to several problems that can be grouped together under the label `the measurement problem':
What selects a preferred basis for physical quantities in nature;
why are interference effects not observed for macroscopic objects;
what process constitutes a `measurement'
and---most importantly---why do measurements have discrete outcomes \citep{schlosshauer2007}?
The orthodox formulation is unacceptable if we assume that measurements are physical interactions and we seek a fundamental theory that can provide an account of these interactions without reference to an ill-defined concept such as `measurement.'
The orthodox theory is adequate only in that it provides a means
of determining the results of experiments
but not if we wish to have a realist description of the physical world. 

When a quantum system interacts with the environment, quantum coherence is
delocalized into the entangled environment-system state
and we effectively lose the ability to observe it \citep{zeh1970, zurek1981}.
This process, known as decoherence, is effectively irreversible and helps explain the non-observation of macroscopic quantum interference effects, as well as often providing a preferred basis for physical quantities.
However, it does not explain how measurements have unique outcomes \citep[57--60]{schlosshauer2007}.

%When a quantum system interacts with the environment, quantum coherence is delocalised into the entangled %environment-system state and we effectively lose the ability to observe it.
%This process, known as decoherence, is effectively irreversible and helps explain the non-observation of %macroscopic quantum interference effects, as well as often providing a preferred basis for physical quantities.
%However, it does not explain how measurements have outcomes at all, or what chooses a particular outcome from %among the different possibilities allowed by the wavefunction.
%It is the problem of outcomes that is the main issue of the measurement problem.

One way of viewing the measurement problem is to say that
the following three properties are incompatible \citep{maudlin1995}:
\begin{enumerate}
	\item A measurement always give rise to a single determinate outcome.
	\item The wavefunction provides a complete description of all the physical properties of a system.
	\item The wavefunction always evolves according to the Schr\"{o}dinger equation.
	%\item A sysytem possess a determinate value of an observable if and only if the system is in an eigenstate of that observable.
\end{enumerate}

Various possible solutions to the measurement problem have been proposed.
These can be loosely placed into three groups depending on which of the arms of the above trilemma are discarded.
Relative state/many worlds interpretations \citep{everett1957,dewitt1973} give up the first,
hidden variable theories,
such as the de Broglie-Bohm pilot wave model \citep{bohm1952},
opt out of the second,
while physical collapse models, such as our own, reject the third.

Physical collapse theories modify the unitary dynamics of the wavefunction
to make wavefunction collapse a physical process.
The Schr\"{o}dinger equation then becomes an approximation,
applicable at the microscopic level but breaking down at larger scales.
In principle, this difference is testable since it will produce small deviations
from the predictions of orthodox quantum mechanics
(see Sect.\ \ref{Sec:Constraints}).
The task is to find a theory that is well defined at the microphysical level
and that closely approximates collapse to the basis preferred by decoherence.
Such collapse must be stochastic so that it conforms to the Born rule for probabilities.
Any collapse theory must cause superpositions
to collapse to probabilistic mixtures in a time sufficient small to avoid
the observation of macroscopic superpositions
and without having an appreciable effect on states that are resistant to decoherence.

\citet{wigner1967} championed the idea that human consciousness triggers a collapse of the wavefunction.
That is, conscious states resist being forced into superpositions.
Modern theories of consciousness-induced collapse,
where superpositions of conscious states are dynamically suppressed,
are explored by \citet{okon2020} and \citet{chalmers2022}.

Gravitational measures may induce collapse in macroscopic superpositions of different mass density states \citep{diosi1989},
an idea championed by Roger \citet{penrose1996, penrose2014}.
When the structure of spacetime evolves into superpositions over a certain threshold, these superpositions collapse into a definite structure.
Another suggestion is that absorption or emission of photons may trigger collapse
to a photon number state and thereby induce localization of macroscopic objects \citep{simpson2011}.

%Spontaneous collapse could contribute to a solution of a number of cosmological puzzles,
%such as the appearance of inhomogeneity in the early universe from an initial homogeneous state,
%the black hole information paradox
%and the origin of the second law of thermodynamics \citep{okon2017}.

The first mathematically detailed spontaneous collapse model to be discussed in the literature
was the theory of Ghirardi, Rimini and Weber \citep{grw1986, grw1988}
%(see Section~\ref{Sec:ExistCollapse})
though later efforts concentrated on the similar
Continuous Spontaneous Localization model \citep{gpr1990}.
Details of the GRW model are presented in the next section.
Comparison between our model and GRW/CSL models follows in Sect.\ \ref{Sec:Advantage}.

\section{Existing Spontaneous Collapse Theories}
\label{Sec:ExistCollapse}
The general idea behind the GRW model is that every particle in the universe is subject,
at random times, to approximate spatial localizations.
The effect of a localization is to cause the wavefunction representing the particle to collapse instantaneously.
The following discussion is based on that in \citet[201--209]{bell1987}.

Suppose that, before collapse, a particle is part of a system represented by a wavefunction of $N$ particles:
\begin{equation}
	\Psi (\x_{1}, \x_{2}, \ldots, \x_{N} ,t).
\end{equation}
The effect on the wavefunction of a localization of particle $k$,
where $k$ is randomly chosen from the $N$ particles,
is produced by multiplying the wavefunction by a single-particle `jump factor' $j(\x' - \x_k)$,
where $\x'$ is a specific value chosen according to the probability density given
by Eq.\ (\ref{Eq:ProbDensity}) below.
%and $\x_k$ is randomly chosen from the arguments of the wavefunction.
The wavefunction is at the same time normalized to unity.
%so that the probability density in (\ref{Eq:ProbDensity}),
%integrated over all space equals one.
That is, the new wavefunction is
\begin{equation}
	\label{Eq:Normalization}
	\Psi'(\x_{1} , \ldots , \x_{N} ,t) = {\frac{j(\x' - \x_{k} )\Psi (\x_{1}, \ldots , \x_{N} ,t)}{ \parallel j(\x' - \x_{k} )\Psi (\x_{1}, \ldots , \x_{N} ,t) \parallel }}.
\end{equation}
%where 
%\begin{equation}
%\Phi (\x_{1} ,\ldots , \x_{N} ,t) = j(\x' - \x_{k} )\Psi (\x_{1}, \ldots , \x_{N} ,t).
%\end{equation}
GRW suggest that jump factor $j(\x' - \x_k)$ is a Gaussian of the form
\begin{equation}
	\label{Eq:jump}
	j(\x' - \x_{k}) =  \left( \frac{\alpha}{\pi} \right)^{3/4}\, {\rm \exp } \left( \frac{-\alpha(\x'-\x_{k})^{2}}{2} \right),
\end{equation}
where $\alpha$ is a new constant of nature.
The effect of multiplying by the Gaussian is to approximately localize the particle within a radius of $1/\sqrt{\alpha}$.
The probability density of the Gaussian being centred at point
$\x'$ is taken to be  
\begin{equation}
	\label{Eq:ProbDensity}
	P(\x', t) =   \parallel\Psi'(\x_{1}, \ldots , \x_{N} ,t) \parallel^{2}  = \int d\x_{1} \ldots d\x_{N} \, |\Psi'(\x_{1}, \ldots, \x_{N}, t)|^{2}.
\end{equation}
This ensures that the probability of collapse is greatest where the magnitude of the wavefunction is greatest, in close agreement with the probabilistic predictions of standard quantum mechanics.
Localizations for each particle occur at random times with mean frequency $\lambda$,
and between each localization the wavefunction evolves according to the Schr\"{o}dinger equation.
By choosing appropriate values for $\alpha$ and $\lambda$,
the theory attempts to make collapses for a microscopic system very infrequent
while collapses for a macroscopic system occur frequently,
making superpositions of such systems unobservable.
The values normally chosen are $1/\sqrt{\alpha} = 10^{-7} {\rm m}$  and $\lambda = 10^{-16} {\rm s}$ \citep{grw1986}
meaning that, for a single particle,
a collapse would occur on average every $10^{9}$ years while for a macroscopic collection of, for example,
$10^{23}$ particles,
a collapse would occur every $10^{-7}$ seconds.

Since collapses will be common for macroscopic systems,
they will be prevented from entering into superpositions of macroscopically separate locations,
thus providing a solution to the measurement problem for macroscopic systems
while maintaining normal continuous (Schr\"{o}dinger) evolution for microscopic systems.

A quantum mechanical state can be represented in many mathematically equivalent ways,
the representation depending on which observables are used to define the basis states of the representation.
The theory of GRW gives a privileged place to the position observable, since the localizations occur
in position space rather than momentum space, or the spaces of other observables.
%The new theory presented later similarly privileges the position representation.

One problem with the GRW model is that,
for wavefunctions of systems containing identical particles,
the existing symmetry of the wavefunction is destroyed when the wavefunction collapses.
The symmetry (or antisymmetry) of the wavefunction with respect to the exchange of identical particles is a requirement of standard quantum mechanics.
The problem arises in the GRW model since, when a collapse occurs,
the localization is focused on a single particle, with only small effects on other identical particles that may be represented by the wavefunction.

In part due to the problems with the GRW model,
continuous spontaneous localization models were subsequently developed \citep{gpr1990}.
These theories preserve the symmetry character of the wavefunction.
We will not describe these theories here,
since the new theory of collapse described in the current work more closely resembles the GRW theory,
and CSL does not differ significantly from GRW on the issues that will be central to this paper.
Similarly, we will not consider relativistic versions of the GRW model \citep{tumulka2006a}.

\section{Discrete Physics}
\label{Sec:DisPhys}
Discrete physics is characterised by the quantities representing the state of a system being discrete valued and finite in number \citep{zuse1970, feynman1982, minsky1982, fredkin1990, zenil2013, baraldo-de-araujo2017}.
%There is speculation that the whole of physical reality is digital:
This is the approach taken by those modelling physics using cellular automata \citep{wolfram1986, wolfram2002, hooft2016}.
In a cellular automaton, the state of a physical system is taken to be represented
by a certain finite number of discrete magnitudes,
and these quantities are defined only at points
on a spatial lattice with a finite separation.
The magnitudes are updated at discrete time intervals according to set rules,
representing the fundamental laws of physics.
Wolfram, t'Hooft, and others
take this approach to be a realistic model of physical systems
rather than merely a means of modelling continuous physical phenomena.
%This approach could be a realistic model of a discrete  physics
%or merely a means of modelling continuous physical phenomena.

Aspects of string theory and other quantum theories of gravity have led many authors to suggest
that there may be a minimum length scale in the universe, at approximately the Planck length, $10^{-35} {\rm m}$
(see, for example, \citet{witten1996, baez2003}).
To many, such as \citet{hooft1997, hooft2016, hooft2018},
this suggests that space-time may be discrete rather than continuous.
Discrete space (and time) may also arise from quantum theories of gravity known as loop quantum gravity \citep{rovelli1995, rovelli2008, smolin2001}.
Causal set theory \citep{dowker2006} is another promising approach to quantum gravity, where causality is given primacy.
Spacetime is modelled as a set of causally linked points,
and is necessarily discrete.
%There is speculation that the whole of physical reality is digital
%(see for example, \citet{deutsch1997, lloyd2000, wolfram2002})
%and we shall make that assumption.
%\citet{hooft2001, hooft2016, hooft2018} refers to Fredkins ideas
%on cellular automata as indicating an approach to discrete physics \citep{fredkin1990}.
\citet{beane2014} have considered possible observational consequences of a discrete space-time.

We will adopt the following principle:
discretize physics not with the aim of approximating continuous reality,
but with aim of considering the consequences of reality actually being discrete.
Unlike \citet{hooft2016},
we will not be assuming that the underlying discrete physics is local or linear,
and unlike \citet{wolfram2002}, we will not be assuming that it is reversible.
%In fact, as we shall see, the discrete physics we end up with
%is non-local, non-linear, and irreversible.
%One of the advantages of the theory that we introduce in Sect.~\ref{Sec:CCQM}
%is that it follows naturally from the discretisation of quantum mechanics.
%Accordingly, we will take as our starting point a discrete-valued wavefunction.
Like many of these approaches,
part of the motivation for us in a discrete physics is to facilitate the emergence of space or spacetime
from something more fundamental
\citep[ch.\ 3]{leckey1998}.
This paper investigates the consequences for quantum mechanics of this discretization.
We will be considering the discretization of space, configuration space, and physical magnitudes;
%Time could also be discretized,
we will not be investigating the consequences of time discretization.
Below we describe one way wavefunctions could be represented in a discrete physics.
The discretization of the wavefunction will be carried out in the position representation.

Consider the wavefunction in configuration space for a system of $N$ interacting particles,
neglecting spin.
In standard quantum mechanics,
the wavefunction is continuous in magnitude and defined at all points in configuration space.
The wavefunction can be written as a product of a magnitude and phase factor as follows:
\begin{equation}
	\Psi (\x_{1}, \x_{2}, \ldots, \x_{N}, t) = f(\x_{1}, \x_{2}, \ldots, \x_{N} ,t) \, e^{i\theta (\x_{1}, \x_{2}, \ldots, \x_{N}, t)},
\end{equation}
where $f$ and $\theta$ are real-valued functions.
The wavefunction is assumed to be normalized to unity in the usual manner.
In a discrete physics, both position space and configuration space are divided into cells of small finite size. Consider dividing configuration space into cells of volume $a_1^3 \ldots a_N^3$, where $a_k$ is the length characteristic of the $k$th particle represented by the wavefunction.
Here the possibility has been left open that the length of the sides of the cell in configuration space may depend on the mass or energy of the particle to which they correspond.
If the length turns out to be independent of the particle and its energy then the volume of a single cell will be $a^{3N}$.
We favour the option that $a_k$ is approximately proportional to the mean de Broglie wavelength of that particle
since this wavelength roughly characterises the rate of change of the wavefunction with distance
%Further discussion of this point is beyond the scope of this paper but more details are given in
%[removed to maintain anonymity].
\citep{leckey1998}.
We define a wavefunction that is both
spatially discrete, with one value per cell in configuration space, and discrete valued:
%That is, take $f$ and $\theta$ as discrete:
\begin{eqnarray}
	f(\x_{1}, \x_{2}, \ldots , \x_{N} ,t) &=& n_{f}(\x_{1}, \x_{2}, \ldots, \x_{N} ,t) \, f_{0}\, \\
	\theta (\x_{1}, \x_{2}, \ldots, \x_{N}, t) &=& n_{\theta}(\x_{1}, \x_{2}, \ldots, \x_{N}, t)\, \theta _{0}\,,
\end{eqnarray}
where $n_{f}$ and $n_{\theta}$ are functions that yield natural numbers,
$f_{0}$ is the base magnitude of the discrete wavefunction and $\theta_{0}$ is the base phase angle.

%We can  a continuous-valued (but spatially discrete) wavefunction by specifying that whenever $|\Psi|$ falls below
%$f_0$, the discrete wavefunction has magnitude $0$.
%In general, the discrete-valued wavefunction has magnitude
%$n \, f_0$ wherever the continuous-valued wavefunction has a magnitude between $n \, f_0$ and $(n+1) \, f_0$.
Define the relative volume of the wavefunction, $v$ as the $3N$-dimensional volume in configuration space, $V$
for which the wavefunction is non-zero, divided by the corresponding volume
of a single cell in $3N$ dimensions:
\begin{equation}
	v = \frac{V}{a_{1}^{3} \ldots a_{N}^{3} }.
	%v = \int_{V} d\x_{1} \ldots  d\x_{N} / \int_{0}^{a_{1}} \cdots \int_{0}^{a_{N}} d\x_{1} \ldots  d\x_{N} = \frac{V}{a_{1}^{3} \ldots a_{N}^{3} }.
\end{equation}
The discrete wavefunction has a clear boundary, unlike the case in orthodox quantum mechanics
where all wavefunctions extend over the entire universe,
since the wavefunction only goes to zero over a finite region
where the potential is infinite, and this is not physical.

If we have $N$ spin-half particles,
the full wavefunction will be given by multiplying the spatial wavefunction
by a spin vector of $2^{N}$ components,
so that the total wavefunction is given by $2^{N} \, v$ complex values.
We argue that the relative volume is a possible measure of the {\em complexity} of the wavefunction,
since the number of complex values required to specify a wavefunction of $N$ particles is proportional to $v$.

\section{Critical Complexity Quantum Mechanics}
\label{Sec:CCQM}
We introduce a modified quantum mechanics that not only provides a new approach to solving the measurement problem by providing a mechanism for wavefunction collapse but admits a realist interpretation of wavefunctions that represent less than the entire universe.

In standard quantum mechanics,
when we write a wavefunction of a single particle or small group of particles,
this is an approximation that we obtain by neglecting the (small) interaction with the environment.
Under many interpretations of quantum mechanics,
including the orthodox one,
wavefunctions are regarded as merely convenient constructions useful for making calculations.
Under this anti-realist interpretation of wavefunctions,
under no circumstances will the wavefunction be regarded as corresponding to what exists in nature.

The theory that we put forward here differs considerably from standard quantum mechanics
in providing for the existence of wavefunctions that represent a finite number of particles,
much smaller than the total number of particles in the universe,
and in allowing a realist interpretation of these wavefunctions and of wavefunction collapse.
%In what follows, we will adopt a realist interpretation since this fits well with the proposed theory.
Furthermore, under this theory the number of particles a wavefunction represents can change with time, either through the wavefunction splitting into two or more smaller wavefunctions, or by combining with one or more other wavefunctions.
%Although the number of particles represented by a wavefunction may vary with time,
%this number will always be considerably less than the total number of particles in the universe.
From this point of view it is perhaps natural that a collapse or split of a wavefunction be triggered when it reaches some critical size, or complexity, for some measure of the complexity of the wavefunction. This takes up a suggestion by \citet[p.\ 598]{leggett1984} that there may be ``corrections to linear quantum mechanics which are functions, in some sense or other, of the {\em degree of complexity} of the physical system described.''

We propose the measure of complexity of a wavefunction (for the moment neglecting spin)
is given by its relative volume in configuration space, as defined in the previous section.
Configuration space has $3N$ dimensions for a system of $N$ particles
so that the dimensionality of a `volume' in this space
will vary with the number of particles represented by the wavefunction.
However, the relative volume is the number of cells occupied by the wavefunction
and so is a dimensionless quantity.
A fuller discussion of the relation between relative volume and complexity,
and a possible link between relative volume, entropy and the arrow of time,
is given in
%[removed to maintain anonymity].
\citet{leckey1998, leckey2016}.

We assume that the wavefunction of a quantum system will collapse, or split,
when it reaches a certain critical relative volume in configuration space.
What a `collapse or split' of the wavefunction amounts to is explained in the following sections.
This modified quantum mechanics that involves an altered dynamics associated with the critical complexity of wavefunctions
we label Critical Complexity Quantum Mechanics (CCQM).
%The collapse or split is a non-linear mechanism akin to that invoked in standard quantum mechanics by the process of measurement.
%However, in CCQM the collapse is triggered by a well defined physical mechanism independent of the presence of observation or measurement,
%thus providing a mechanism for solving the measurement problem.

\section{Wavefunction Collapse in CCQM}
\label{Sec:Collapse}
We propose that there exists a critical relative volume in configuration space
such that, when the wavefunction reaches this volume, it will undergo a non-linear collapse,
resulting in the relative volume reducing below
the critical relative volume.

The existence of a critical volume is clearly motivated in the case
where the wavefunction is in reality discrete valued.
%The relative volume of the wavefunction in configuration space determines
%the average wavefunction magnitude because the wavefunction remains normalised at all times.
%(Note: it is to be understood that in a fully discrete physics
%where integrals are used they should, strictly speaking,
%be replaced by sums.)
Thus, if a wavefunction represents the state of a large number of particles
having probability distributions spreading out in position space,
then as the relative volume increases with time,
eventually the wavefunction magnitude would fall everywhere below $f_0$.
Hence, the discrete wavefunction would become zero everywhere in configuration space,
meaning that the system would effectively cease to exist!
%\footnote{
	%If we were to assume that the entire wavefunction were to be renormalised every time a cell magnitude fell below $f_0$,
	%then rather than disappearing altogether, the wavefunction would vanish in localized regions instead,
	%leading to the local fragmentation of the wavefunction.
	%Local deviations from the predictions of quantum mechanics would occur,
	%of a kind that we do not observe,
	%motivating the type of collapse discussed in this section.}
\citet[p.\ 537]{minsky1982} discusses the problem of how to represent spherically propagating waves in position space in a discrete physics,
and recognises that the tendency of the magnitude to everywhere
fall below the threshold magnitude represents a problem.
He does not, however, suggest a solution.
It can be seen that a similar problem arises for discrete physics
in the case of a wavefunction spreading in a $3N$ dimensional configuration space.
%Since the magnitude of  the wavefunction is discretized,
%exact nomalization cannot be maintained as the wavefunction spreads.
%Consider a toy example where a particle occupying a single cell with magnitude ten spreads to two cells with magnitude seven each. In so doing, the wavefunction increases in volume in configuration space, as required in our model but deviates from normalisation by 2%.
%However, in a realistic system with a large volume in configuration space,
%approximate nomalization is maintained to a high precision.
%This would not have physical consequences since normalisation only has an affect when the wavefunction collapses,
%and at this point the wavefunction undergoes normalisation.

We suggest that a natural solution to the problem of spreading wavefunctions in discrete quantum mechanics is for the wavefunction to `collapse' to some extent when it reaches some critical relative volume instantaneously (or virtually instantaneously) reducing in volume.
Alternatively, the wavefunction could split into two or more smaller wavefunctions;
this possibility will be discussed in Sect.\ \ref{Sec:NonLinear}.
We suppose, like GRW, that the wavefunction will be normalized to unity after every collapse
(see Eq.~(\ref{Eq:Normalization}) with integrals replaced by the appropriate summations)
so the average magnitude per cell will rise when a collapse occurs.

The wavefunction would then resume spreading until it again reached the critical volume,
at which time it would collapse again, and so on.
It appears that some degree of spontaneous localization of wavefunctions
is a natural consequence of a discrete physics.

%In this discussion,
%we have assumed that the corresponding continuous wavefunction is normalized to unity at all times
Schr\"{o}dinger dynamics maintains the normalization of the corresponding continuous wavefunction at all times,
and the discretized wavefunction approximates this wavefunction.
%If we were to assume instead that the discrete wavefunction is to be normalized as closely as possible to unity at all times,
%then rather than disappearing altogether when the continuous wavefunction dropped below $f_0$ everywhere,
As the continuous wavefunction drops below $f_0$ in regions of configuration space,
the discretized wavefunction would vanish in localized regions,
leading to the local fragmentation of the wavefunction.
Local deviations from the predictions of quantum mechanics would occur, of a kind that we do not observe.
The form of collapses adopted in this paper are designed not only to maintain
the average magnitude of the wavefunction above a certain minimum level,
but also to prevent the fragmentation of the wavefunction,
by localising it in a single contiguous region.
Assuming that $f_0$ is constant,
the normalization of the discretized wavefunction will not in general be exact,
but given that the relative volume is always very large,
approximate normalization is maintained to a high precision.
The probability distribution of the location of the collapse centre
is given by a discretized version of Eq.~(\ref{Eq:ProbDensity}) which is normalized,
so the probabilities will add to one.
%The probabilities of the centres of collapse are given, just in GRW, by the ratio of the magnitude at a certain point to the whole,
%and it is these probabilities that will determine the empirical predictions of the theory.
%The deviation from exact normalisation may have empirical consequences, but
%in under either way of normalising,
%the  normalisation is restored as closely as possible when a collapse occurs.

For this solution to be viable, the collapse must not cause deviations from standard quantum mechanics that are in conflict with the results of experiments that have already been carried out. Consider the wavefunction of a single particle in free space.
In reality, single-particle wavefunctions may be rare,
but suppose for the moment that weakly interacting particles are in reality represented for much of the time by wavefunctions
of only one or a small number of particles.
%(This supposition is discussed in more detail in the next section.)
We can use these wavefunctions to put some constraints on the size of the critical relative volume, $v_c$.
We know, from interference experiments, that a wavefunction of a free particle can spread over large volumes in position space before two parts of that wavefunction interact at a detector to produce an interference pattern.
If the critical volume $V_c$ is not large enough then such wavefunctions would collapse before reaching the detector,
thus destroying the interference effects and leading to conflict with experimental results.
This puts a lower bound on the value of the critical volume and hence on the critical relative volume $v_c$.
The volume that the single-particle wavefunction reduces to after collapse must also be sufficiently large otherwise the collapse would have observable consequences.
Some of these consequences are considered in Sec.~\ref{Sec:Constraints}.

It should be noted that an assumption is being made that the limitation
on the relative volume will apply to photons as well as particles with mass.
While there are certain difficulties associated with assigning a wavefunction to a photon,
it has been demonstrated that this can be done as long as the wavefunction is interpreted slightly differently than is usual in elementary quantum mechanics.
Methods of defining a photon wavefunction are given by \citet{bialynicki1994, bialynicki2020} and \citet{sipe1995}.
According to these definitions, the probability interpretation and normalization condition for the photon
wavefunction differ slightly from the usual ones,
but not in a way that significantly affects the arguments in this paper.
The full treatment of photons is beyond the scope of this paper.
Since photons are relativistic particles,
they must be treated within an extension of CCQM to relativistic quantum theory.

\section{Solution to the Measurement Problem}
\label{Sec:SolnMeasProb}
%The type of collapse proposed in the previous section for a few-particle wavefunction is not itself a solution to the measurement problem. 
We have not yet discussed localizations to small regions,
such as spots forming on photographic plates, that take place when measurements occur.

Consider an $N$-particle system in which each of the particles interacts relatively strongly with every other particle in the system.
In our model, such a system is likely to be represented by a single wavefunction.
Suppose that the position probability distribution of each particle in the $N$-particle interacting system covers a volume of $V_s$
%(or a relative volume of $v_s = V_s/a^3$)
in position space.
Then the corresponding relative volume of the wavefunction in configuration space will be of the order $v_s^N$,
since the wavefunction will be spread over a distance of order $V_s^{\frac{1}{3}}$
in each orthogonal direction in the $3N$ dimensional space.
Thus the relative volume will tend to increase exponentially with $N$.
%the number of particles represented by the wavefunction.
Here the assumption has been made that the length of the cell-side, $a$ is the same for each particle,
which is a reasonable approximation for an order of magnitude calculation.

As a result, as $N$ grows, the relative volume of the wavefunction will grow very rapidly, and will readily reach the critical relative volume $v_c$.
For an interacting, many particle system, a small spread in volume of the probability distribution for each particle in position space will contribute a large amount to the relative volume of the wavefunction in configuration space.
Before the position probability distribution of each particle spreads to a macroscopic volume,
the critical relative volume for the wavefunction will be reached, bringing about a collapse,
thereby restricting each particle to a smaller volume in position space.

Thus, many-particle interacting systems will have particles within them that tend to remain localized in position space, consistent with our observations.
Furthermore, collapse of the wavefunction in this manner will prevent superpositions of large numbers of interacting particles spreading over macroscopic volumes in position space.
In this way, the collapse mechanism will prevent the occurrence of macroscopically distinguishable superpositions.
%that would otherwise arise in the application of the unmodified Schr\"{o}dinger equation to a typical measurement process,
%where no collapse is assumed.
Instead, the wavefunction will collapse and give rise to a determinate outcome to the measurement.
Conversely, a system of a single particle, or a small number of interacting particles, can spread over larger volumes in position space before the wavefunction representing the system reaches the critical relative volume and collapses, permitting the interference effects that we observe.
%Thus we claim that the critical complexity theory of the collapse of the wavefunction can provide a solution to the
%measurement problem.
Thus, we claim that the CCQM theory of wavefunction collapse
explains why superposition effects sometimes occur,
even over large distances, and yet we never see superposition effects for macroscopic objects.
This provides a solution to the measurement problem.

\section{Proposed Collapse Process}
\label{Sec:PropCollMech}
To keep the wavefunction from disappearing below the minimum threshold representable in a discrete physics
requires a reduction in size of the wavefunction by
splitting into two or more separate wavefunctions or by a `collapse',
localising in configuration space by
reducing in volume to some fraction, $F, \; 0<F<1$ of the original volume.
For simplicity, we suggest a fraction of one-half.
The actual fraction is not important as long as the volume remains large after the collapse.
We will now make a tentative suggestion about the form this collapse might take by adapting a simple system of collapse from the GRW model.
We label this model the `jump' model of collapse.
In this model, when a collapse occurs, the original wavefunction $\Psi(x, t)$, of $N$ particles,
is multiplied by a `jump factor' $j(x'-x)$ where $x \equiv (\x_1, \ldots, \x_N)$ and $x'$ is the centre of the collapse in configuration space.
As in the GRW model, the jump factor $j(x)$ is a Gaussian, but here a Gaussian in configuration space rather than a Gaussian in position space; $j(x)$ is a product of single-particle Gaussians:
\begin{equation}
	j(\x_{1}, \ldots, \x_{N} ) = j(\x_{1} )\ldots  j(\x_{N} ),
\end{equation}
where 
\begin{equation}
	j(\x_{i}) =  \left( \frac{\varepsilon}{\pi} \right)^{3/4}\, {\rm \exp } \left(- \frac{\varepsilon \x_{i}^{2}}{2} \right) \, .
\end{equation}
%and is symmetric under the interchange of any two particles.
Since we are dealing with discrete physics,
the Gaussians would be discretized,
dropping to zero at the point where the magnitude drops below $f_{0}$.
%The volume of the wavefunction in configuration space is thus reduced.
Just as in GRW, the wavefunction is normalized at the collapse
and the probability distribution of collapses is derived from the normalized wavefunction
by discretized versions of Eqns.~(\ref{Eq:Normalization}) and (\ref{Eq:ProbDensity}),
as noted in Sect.~\ref{Sec:Collapse}.
The effect of the collapse is to localize every particle in the wavefunction
approximately to within a radius of $1/\sqrt{\varepsilon}$.
The volume of the wavefunction in configuration space will be the region
where the (discrete) wavefunction is nonzero,
and this will be reduced by multiplying by the Gaussian.
%The actual spread of each particle will depend of how quickly the wavefunction goes to zero
%\citep[Appendix B]{leckey2023}.
The value of $\varepsilon$ will vary in each
particular case, being the
value required to reduce the relative volume of the wavefunction by the fraction $F$.
%$\varepsilon$ will depend on the constants $F,\, v_c,\, f_0$, and $a_k$.
The value of $\varepsilon$ will be the same for each particle in the wavefunction
so the original symmetry of the wavefunction will be maintained.
We are not considering in detail experimental constraints in this paper,
so we will not attempt to put rigorous bounds on the constants of the theory;
however, some observational constraints on the theory will be discussed in Sect.~\ref{Sec:Constraints}.
%and in the appendices.
%We assume the probability distribution of the collapse centre and renormalisation of the %wavefunction after collapse would be determined in the same way as in the GRW model, so as to %closely preserve the statistics of the wavefunction.
The important point to note is that the value of $\varepsilon$ will be much smaller for systems of a few particles than for systems of a large number of particles,
as discussed in Sect.\ \ref{Sec:SolnMeasProb}.

\section{A New Non-linear Dynamics}
\label{Sec:NonLinear}
The modified quantum mechanics of CCQM must involve further changes to linear quantum mechanics
other than simply introducing wavefunction collapses.
%which localise the wavefunctions in configuration space.
%The model must also involve alterations to the dynamics of linear quantum mechanics
%that have the effect that the number of particles represented
The model must have processes which have the effect that the number of particles represented by
by a single wavefunction remains finite and changes with time.

Outside collapse or spitting/combining events,
Schr\"{o}dinger dynamics is maintained.
This would be a discretized version of the Schr\"{o}dinger equation
that the continuous version closely approximates.
%The details of this dynamics are beyond the scope of the paper.
%\citet[p.\ 248]{hooft2016} suggests replacing the derivatives with difference equations,
%which he calls lattice derivatives,
%whereby the evolution of a cell value with time depends only on immediately neighbouring cells,
%but he does not develop the idea in detail.
The Schr\"{o}dinger equation can be discretized in space and time in various ways;
for recent discussions, see \citet{tarasov2016} or \citet{chou2021}.
%We discuss only the processes of combining, splitting and collapse.

In this modified dynamics, when a system described by a spreading wavefunction
begins to overlap with another wavefunction,
there is a certain probability, depending on the amount of overlap and the strength of interaction between the particles of the two systems,
that these systems will combine into a single wavefunction.
%In this modified dynamics, a system described by a single, spreading wavefunction,
%when it comes into contact with another wavefunction, is postulated to evolve as follows.
%Suppose that the position probability distributions of the particles in the first wavefunction evolve
%in such a way that they increasingly overlap with those of particles represented by the second wavefunction.
%Then there is a certain probability,
%depending on the amount of overlap and the strength of interaction between the particles of the two systems,
%that these systems will combine into a single wavefunction.
The simplest way to combine normalized wavefunctions $\Psi_a$ and $\Psi_b$,
of $N_{a}$ and $N_{b}$ particles, respectively,
is to replace them with the symmetrized product wavefunction:
\begin{equation}
	\Psi_{ab} (\x_{1}, \ldots, \x_{N} ,t) = S \, \Psi_{a} (\x_{1}, \ldots, \x_{N_{a}}, t) \otimes \Psi_{b} (\x_{N_{a}+1}, \ldots, \x_{N} ,t).
\end{equation}
where $N = N_{a} + N_{b}$ and $S$ is the normalized symmetrization operator which ensures that the combined wavefunction
is symmetric with respect to the exchange of identical bosons,
and antisymmetric with respect to the exchange of identical fermions.

It is often considered that all identical fermions must be antisymmetrized
and all identical bosons symmetrized under exchange of particles
without exception \citep[p.\ 204]{griffith2005}.
However, as noted by \citet[p.\ 570]{french1978},
when the amount of overlap of the wavefunctions of two systems is negligible,
there are no observable consequences of representing the systems
in separate wavefunctions without symmetrisation.
In CCQM, there may be some empirical deviations from standard quantum mechanics
resulting from the small overlap of separate wavefunctions without symmetrisation.
%\footnote{
	%	It might be thought that because non-interacting systems are routinely represented by product wavefunctions in standard quantum mechanics,
	%	the `change' from separate wavefunctions to a product wavefunction represents no real change at all.
	%	After all, the product and the separate wavefunctions give rise to the same empirical predictions in standard quantum mechanics.
	%	However, it should be remembered that what is sought is a realist interpretation of the wavefunction,
	%	and under a realist interpretation a single wavefunction in a $3N$ dimensional configuration space
	%	is quite a different entity to a pair of wavefunctions in lesser dimensional configuration spaces,
	%	even if the former is equal to a product of the latter.
	%}

It will generally be the case that the combined wavefunction will exceed the critical relative volume
and immediately undergo a collapse
of the kind given in Sec.~\ref{Sec:PropCollMech}.

We suggest that the probability per unit time of combining wavefunctions $a$ and $b$,
at time $t$, goes like:
\begin{equation}
	\label{Eq:ProbCombine}
	\mathcal{P}_{ab}^{\, c}(t) \sim \gamma_{ab} \, \Theta_{ab}(t) \, ,
\end{equation}
where $\gamma_{ab}$ is a dimensionless measure of the strength of interaction between
the particles in the two wavefunctions,
and $\Theta_{ab}(t)$ measures the  overlap of the wavefunctions at time $t$ in $N$ dimensional configuration space:
\begin{equation}
	\Theta_{ab}(t) = \int \ldots \int\Psi_{a}^{*}(\x_{1}, \ldots , \x_{N_{a}}, t)
	\Psi_{b}(\x_{N_{a}+1}, \ldots, \x_{N}, t) \, d\x_{1} \dots d\x_{N} \, .
\end{equation}
One option for $\gamma_{ab}$ would be to average the strengths of interaction
between the particles of one wavefunction with the particles of the other.
%between every pair of particles, one in each wavefunction:
%\begin{equation}
%	\gamma_{ab} = \frac{1}{N_{a} + N_{b}} \, \sum_{i=1}^{N_{a}} \sum_{j=1}^{N_{b}} \tilde{\gamma}_{ij},
%\end{equation}
%where $\tilde{\gamma}_{ij}$ represents the strength of interaction
%where $\tilde{\gamma}_{ij}$ is a dimensionless measure of the strength of interaction
%between particle $i$ in wavefunction $a$ and particle $j$ in wavefunction $b$.
%The combined wavefunction formed this way will most likely have a volume over the critical volume,
%so it will collapse after combining.

When a wavefunction reaches the critical relative volume, there is some probability that the wavefunction splits.
%Firstly, we identify the range of wavefunctions into which the wavefunction $\Psi_{c}$ can split
%such that the split wavefunctions are together empirically indistinguishable from $\Psi_{c}$ at the %time of the split.
%In order to do this, it is helpful to identify the probability density of particle $k$ in position space, $g(\x_{k})$:
The pair of split wavefunctions should be empirically indistinguishable from the original wavefunction, $\Psi_c$,
at the time of the split.
First, introduce the probability density of particle $k$ in position space:
\begin{equation}
	\label{Eq:PosnProb}
	g(\x_{k} ) =  \int \ldots \int | \Psi (\x_{1}, \ldots, \x_{N}, t)|^{2} d\x_{1} \ldots d\x_{k-1} d\x_{k+1}, \ldots  d\x_{N}.
\end{equation}
That is,
$g(\x_k) d\x_k$ is the probability of finding, upon measurement of position,
particle $k$ within volume $d\x_k$ of $\x_k$ in position space.
Now, consider an equivalence class of unsymmetrized wavefunctions $[\Psi_{c}]$
such that any member, $\Psi'_c$, of the class preserves the probability distribution of identical particles in the original wavefunction $\Psi_{c}$.
That is, suppose that there are $n_{j}$ particles of kind $j$ in $\Psi_{c}$,
labelled $j_{1}$ to $j_{n_j}$.
Then the average position probability density of the $j$ particles,
\begin{equation}
	\bar{g}(\x_j) = (g(\x_{j_1}) + \ldots + g(\x_{j_{n_j}}))/n_{j},
\end{equation}
is the same in $\Psi'_{c}$ as in $\Psi_{c}$ for each kind of particle $j$.
The wavefunction $\Psi_{c}$ of $N = N_{e} +  N_{f}$ particles can split into wavefunctions $\Psi_{e}$ and $\Psi_{f}$ of $N_{e}$ and $N_{f}$ particles, respectively,
with the new wavefunctions determined from some unsymmetrized wavefunction $\Psi'_{c}$ by integrating out the position coordinates
of the particles not represented in the respective wavefunction:
\begin{equation}
	\Psi_{e}(\x_{1}, \ldots , \x_{N_{e}}, t) = \int \ldots \int \Psi'_{c}(\x_{1}, \ldots , \x_{N}, t) \, d\x_{N_{e}+1} \ldots d\x_{N},
\end{equation}
and similarly for $\Psi_{f}$.
Immediately after the split,
the $\bar{g}(x_{j})$ are unchanged,
and so the split system is empirically indistinguishable from the original system at that instant.

A possible rule for splitting is as follows.
The probability that $\Psi_c$ splits into $\Psi_e$ and $\Psi_f$
depends on the proximity of the product  $\Psi_{e} \otimes \Psi_{f}$ to the unsymmetrized wavefunction,
$\Psi'_{c}$, as measured by their overlap:
\begin{equation}
	P^{c}_{ef} (t) \sim \Theta_{(e \otimes f) c'} (t),
\end{equation}
where $\Theta$ is evaluated at the time, $t$, that $\Psi_c$ reaches the critical volume.
The idea is that the wavefunction will likely split into subsystems that more closely approximate non-interacting systems.
(A pair of non-interacting systems can be represented by a product of two separate wavefunctions.)
If the wavefunction does not split then it must collapse by localising in configuration space,
so bringing the relative volume below the critical value.

%As noted by \citet[p.\ 570]{french1978},
%when the amount of overlap of the wavefunctions of two particles is negligible,
%there are no observable consequences of representing the particles
%in separate wavefunctions without symmertrisation.
%In CCQM, there may be some overlap of wavefunctions that have just split,
%and this may be empirically observable.

The specific probabilistic rules for the combining and splitting of wavefunctions that we have put forward are meant as suggestions only, and may have to be altered in light of empirical evidence.
%As noted earlier,
%at the instant of splitting there is no observable difference between the split wavefunctions
%and the original.
%However, as the wavefunctions evolve in time, they may appreciably overlap and this could lead to observable consequences.
These rules are examples of the kind of rules that will lead to strongly interacting particles
being in wavefunctions of more particles
that collapse more often,
and hence lead to a solution of the measurement problem, as explained earlier.
%How the interaction between particles in separate wavefunctions is handled
%must await an extension of the theory to quantum field theory.

It would seem to follow from the combining rule, Eq.~(\ref{Eq:ProbCombine}),
that non-interacting particles,
such as a collection of photons, cannot combine into a single wavefunction.
However, a system of weakly interacting particles,
such as the photons in black body radiation or particles in a Bose-Einstein condensate,
form by interaction with their environment,
and hence in CCQM can be part of a single wavefunction.

A macroscopic object may consist of many separate wavefunctions that are continually in the process of combining,
splitting, expanding and collapsing.
There will also be frequent exchanges of particles between wavefunctions.
While describing the interaction between particles in separate wavefunctions is beyond the scope of this paper,
the occurrence of the exchange of particles in CCQM suggests that there may be a natural extension to quantum field theory,
where the interaction between systems is described by the exchange of so-called `virtual' particles.

\section{Comparison of CCQM with GRW/CSL}
\label{Sec:Advantage}
One criticism that can be levelled at the GRW/CSL theories is that they are ad hoc---the only motivation
for the modification of linear quantum mechanics is to produce a solution to the measurement problem.
We follow a notion of ad hoc akin to that given by \citet{schindler2018},
whereby a hypothesis modifying a theory is ad hoc if there is no background theory
that would give us reason to believe the hypothesis.  
In our case, discrete physics is the background theory
which motivates the modifications to quantum mechanics introduced by CCQM.
This is an independent motivation for the theory,
and it opens up a new avenue of research into a modified quantum mechanics
that we think is worth pursuing.
%Although it adds some complexity,
%one strength of the CCQM is that it follows naturally from a fully discrete physics.
%The collapse of the wavefunction is well motivated if we consider all of physics to be discrete,
%and is not a mere ad hoc modification to linear quantum mechanics.
%CCQM, on the other hand, is independently motivated, as we have argued above, on the basis of a fully discrete physics.
CCQM also has the potential to satisfy the `rule of simulation' proposed by \citet{feynman1982},
who suggests as a heuristic for discovering physical laws
that we only accept laws that could be simulated on finite digital computers.
For a further discussion, see
%[removed to maintain anonymity].
\citet[p.\ 59]{leckey1998}.
%We hope to discuss this and other features of CCQM in future papers.

The GRW model produces collapse by multiplying the wavefunction by a Gaussian in the three spatial coordinates of one of the particles.
Since the tails of a Gaussian extend to infinity, this only ever approximately localizes the particle.
If we accept that a particle is localized within some region only if the wavefunction goes to zero outside that region, then GRW does not achieve such localization.
If we consider the pointer of a measuring instrument that is in a superposition of possible outcomes prior to a measurement,
then a GRW collapse does not bring about a reduction to a proper mixture of states but merely the same pre-measurement superposition with all but one component dramatically reduced.
These are two aspects of the so-called tails problem;
\citet{mcqueen2015} identifies four variants of this problem in dynamical collapse models.
Given that physical collapse theories take a realist view towards the wavefunction,
we should not ignore the dynamical evolution of the tails.
%One approach is to say that collapse produces one dominant state `for all practical purposes'
%but acknowledge that the universe is in fact rich in a multitude of (extremely small) branches.
%This result can be largely achieved simply by decoherence
%(the Many Decohered Worlds approach, see \citet{cordero1999}).
%Another approach, favoured by \citet{albert1996},
%is to say the universe actually is in the dominant state when the modulus squared
%of its amplitude becomes sufficiently close to one.
%That is, we give up the absolute eigenstate-eigenvalue link of standard quantum mechanics,
%substituting a `fuzzy link',
%and adopting a principle of `vague supervenience' of the localization of particles on wavefunction magnitudes.
%particles can be localised within a finite region and the tails of the wavefunction can be eliminated.
GRW/CSL models fall foul of these tails problems \citep[12--14]{mcqueen2015}. 
%The fourth of McQueen's tails problem is a dilemma:
%if the collapse function is of infinite extent,
%the tails of the wavefunction are not completely eliminated
%while if it has finite support,
%the wavefunction would subsequently expand infinitely fast and may violate relativity \citep{hegerfeldt1998}.
Since CCQM is based on a discrete physics where wavefunction amplitudes become equal to zero once they fall below a certain threshold,
the tails will not be of infinite extent,
and all particles will be localized within a finite region.
McQueen argues that collapse theories that solve the first three tails problem
by a non-Gaussian collapse function that fully localizes particles within a finite region
will result in wavefunctions that spread infinitely fast after collapse,
and this may result in a conflict with relativity.
He calls this the fourth tails problem \citep[p.\ 16]{mcqueen2015}.
%However, this assumes a continuous physics where a  continuous Schr\"{o}dinger equation applies.
In CCQM,
the localization comes about because of 
the discrete nature of the wavefunction and the Gaussian collapse function.
After collapse, the wavefunction will expand with finite speed.
% that uses a disretised Schr\"{o}dinger equation
%--- that the standard continuous one closely approximates ---
%the wavefunction will expand at finite speed after collapse.
%Although the wavefunction is of finite extent,
%it will not expand infinitely fast.
%The continuous equations of motion currently used are approximations to the underlying discrete reality.
%The rate of expansion of the edge of the discrete wavefunction
%is approximated by the rate at which the same point on the corresponding continuous wavefunction would expand,
%and this is finite since the continuous wavefunction is not truncated to a finite volume.
This means that, unlike GRW,
CCQM does not suffer from any of the four variants of the tails problem.

A major point of difference between the theories relates to the treatment of photons.
The original GRW theory proposes that `all constituents of any system' \citep[p.\ 480]{grw1986}
are subject to localizations at the same rate,
but it has been suggested that the localization of photons to the small volume given by the `hitting' Gaussian
(Eq.\ (\ref{Eq:jump})) would produce an effect on the spectrum of the cosmic microwave background radiation
in conflict with observations \citep{squires1992}.
Subsequent versions of the GRW/CSL theory altered this proposal
by making the rate of collapse proportional to the rest mass of the particle \citep{ggb1995},
so that photons are never subject to direct collapse events.
\citet{pearle2018} gives a modified relativistic version of CSL that gives rise to some collapses for photons.
We suggest that it is an advantage of CCQM that the critical volume criterion for collapse can be applied equally to all particles, including free photons.
In Sect.\ \ref{Sec:PropCollMech}, the assumption has been made that the hitting Gaussian
would be extremely large for few-particle systems
such as free photons or neutrinos, so altering the momentum distribution very little,
but by a potentially measurable amount.
It is a virtue of the theory that it leads to precise empirical predictions for astronomical photon
observations that could potentially distinguish CCQM from not only standard quantum mechanics,
but from other collapse models.

A feature of CCQM is that, due to the fact that the relative volume grows exponentially with the number of particles,
there is a rapid transition from the `quantum realm',
where the position probability distributions of particles are free to spread in position space,
to the `classical realm' where their capacity to spread is limited.
In GRW/CSL on the other hand, this transition is linear in the number of particles in a system,
because the rate of localization is linear in the number of particles.
Due to this difference, the transition between the two realms could take place
within the realm of microscopic systems in CCQM,
whereas the transition must take place within the realm of macroscopic systems in GRW/CSL,
which has been noted as a shortcoming by \citet{shimony1989}.

The original GRW model does not preserve the symmetry of the wavefunction.
However, symmetry preserving versions of GRW have been developed \citep{dove1995,tumulka2006b}
where fermions and bosons are localized by separate operators.
%in order to maintain the correct symmetry of the wavefunction.
In our model,
all particles in the wavefunction are localized
by a product of Gaussians that is symmetric under the interchange of any two particles.
Hence, the original symmetry of the wavefunction is maintained.
Although the appropriate symmetry is preserved within a wavefunction 
%Although the wavefunction will be antisymmetric under exchange of any two identical fermions within the wavefunction
there will not be any symmetry under exchange of particles represented by different wavefunctions.
This is an immediate consequence of allowing wavefunctions smaller than the wavefunction of the entire universe.
%However, as pointed out by \citet[p.\ 570]{french1978} this will not give rise
%to observable consequences so long as the separate wavefunctions do not overlap appreciably in position space (see the discussion in footnote 1 of \citet{eisert2020}).
This difference in symmetrisation enables a realist interpretation of a
wavefunction of a finite number of particles.
As noted earlier, this lack of symmetry does not lead to readily observable consequences
where there is little overlap between wavefunctions.

\section{CCQM and Wavefunction Realism}
\label{Sec:WfnReal}
As noted in the introduction, according to standard quantum mechanics,
as long as there is a non-zero strength of interaction between them,
particles must be represented by a single wavefunction,
whose arguments are the positions of all the particles.
It would seem to follow that there can only be a single wavefunction for the whole universe,
and there is nothing in GRW/CSL that alters this situation.
This is to be contrasted with the case of CCQM,
which represents the universe by many wavefunctions, each of which represents a limited number of particles.

According to the standard interpretation of a wavefunction of $N$ particles, the quantity
$|\Psi(\x_1, \ldots, \x_N, t)|^2$
is simply a probability density: $|\Psi(\x_1, \ldots, \x_N, t)|^2 \, d\x_1 \ldots d\x_N$
is the probability of finding, on simultaneous measurement of the positions
of each of the $N$ particles at time $t$,
particle 1 within volume $d\x_1$ of $\x_1$, particle $2$ within $d\x_2$ of $\x_2$, and so on.
This interpretation is adequate if one is only interested in predicting the results of experiments, but can the wavefunction be given a more direct realist interpretation?
CCQM allows for the wavefunction itself to be interpreted realistically
as a complex valued wave (or field) in a $3N$ dimensional configuration.
(Taking spin into account, there are $2N$ complex magnitudes at each point in configuration space: a vector field in configuration space.)
According to \citet[p.\ 80]{ney2021},
``[\ldots] wave function realism is unique in yielding pictures of the world with two intuitively nice metaphysical features: separability and locality.''
Separability and locality apply not in three-dimensional position space
but in $3N$-dimensional configuration space.
For a discussion, see \citet[80--132]{ney2021}.
Unlike other wavefunction realism ontologies,
our model is unique in having wavefunctions that are smaller than the whole Universe
and rules that govern how the number of particles in a wavefunction evolve with time.

This realist interpretation of the wavefunction is consistent with the interpretation suggested by \citet{bell1990}
for the wavefunction in the GRW model:
%that the modifications to quantum mechanics introduced by the GRW model remove the concept of measurement as a primitive and allow
that the magnitude of the wavefunction squared be interpreted as the `density of stuff' of which the world is made;
and this is a density in a $3N$ dimensional configuration space.

%Another quantity of interest is the distribution of each particle $k$ in three-dimensional position space,
%given that this particle is represented in an $N$-particle wavefunction.
%According to the standard probabilistic interpretation of the wavefunction, the quantity $g(x_{k})$ of Eq.~(\ref{Eq:PosnProb)
	%is the probability density of particle $k$ in position space:
	%$g(\x_k) d\x_k$ is the probability of finding, upon measurement of position,
	%particle $k$ within volume $d\x_k$ of $\x_k$ in position space.
	
	We suggest the probability density in position space, $g(\x_k)$, (Eq.~(\ref{Eq:PosnProb})),
	can be interpreted realistically as the density of the particle $k$ in position space: it is a projection of the wavefunction from configuration space into position space.
	In the case of a discrete wavefunction, the volume in position space of a single particle $k$
	can be defined as the volume for which the quantity $g(\x_k)$ is non-zero.
	\citet[52--62]{bell1987} also argues that a realist ontology for quantum mechanics should include what he calls
	`local beables'---objects having approximately determinate locations in $3D$ position space.
	For particles in the wavefunction of a macroscopic object, $g(\x_{k})$ is a local beable.
	
	As discussed in Sect.\ \ref{Sec:SolnMeasProb}, macroscopic systems of strongly interacting particles
	are described by separate wavefunctions, each confined to a localized region,
	so that $g(\x_k)$ is well localized in position space for the particles in these systems.
	This corresponds to the macroscopic world as we are aware of it,
	thus providing an account of the world compatible with our experience.
	
	\citet{ghirardi1996} and \citet{ggb1995} reject the density of stuff
	in $3N$-dimensional configuration space interpretation,
	and instead adopt an interpretation for CSL that involves the average mass density in three-dimensional position space. It is easy to see why the interpretation of $g(\x_k)$ as the density of a particle in position space will not give a picture of reality compatible with experience in the case of CSL for,
	in that model, if we presume that there is one wavefunction for the whole universe and that this wavefunction is appropriately symmetrized under the exchange of identical particles, then every identical particle in the universe must have exactly the same position probability distribution.
	In other words, the distribution $g(\x_k)$ of each identical particle must extend over the entire universe.
	This is not just because the wavefunction is continuous in the GRW/CSL model and so can never fall to zero magnitude. Even if the wavefunction were discrete, the distribution $g(\x_k)$ of each particle
	of the same type as particle $k$ would extend over the entire universe
	(except if there are regions in which particles of this type are absent)
	due to the presumed symmetry character of the wavefunction.
	
	In order to obtain a picture of reality compatible with our experience,
	supporters of GRW/CSL must adopt a particle-number density or mass density interpretation
	of the wavefunction, or a ‘flash’ ontology, in which matter is composed of discrete space-time
	points called ‘events’ or ‘flashes’, one flash for each collapse, located at the collapse
	centre (Bell, 1987; Allori et al., 2008).
	%As Allori et al.\ points out,
	These approaches to GRW are examples of an a priori `primitive ontology' \citep{allori2015}
	which presupposes that the objects of the ontology resides in 3-dimensional position space.
	In CCQM, we do not require this assumption.
	%but we can still derive objects in our ontology
	%that reside in 3D position-space.
	Since identical particles only symmetrize within a single wavefunction
	that represents a finite number of particles extended over a limited volume of space,
	we are able to adopt the density of stuff interpretation of Bell.
	This means that we can be realist about the wavefunction in configuration space
	as well as deriving from it local beables in position space.
	It is an advantage of CCQM that it allows for a more straightforward realist interpretation
	of the wavefunction than is possible in the GRW/CSL models.
	
	\section{Observational Constraints on the Parameters}
	\label{Sec:Constraints}
	A number of new physical quantities are introduced in CCQM.
	Discretizaton of the wavefunction entails the introduction of a minimum magnitude, $f_0$ and minimum phase, $\theta_0$.
	In CCQM, a cell size, $a$ that may be a function of physical parameters such as the De Broglie wavelength of the system,
	and a critical relative volume, $v_c$ are introduced to govern the collapse or splitting of the wavefunction.
	An additional parameter is required to determine the fraction, $F$ by which the wavefunction reduces in size when a collapse occurs.
	
	The CSL and GRW models also contain new physical parameters
	%the localization rate, $\lambda$, which tells us the frequency of %localization (collapse) events,
	%and localization length $1/\sqrt{\alpha}$, which tells us the degree to %which a system is localised.
	%Later versions of these models introduce further parameters
	%that give the mass dependence of the collapse frequency.
	whose values were initially chosen to provide a solution to the measurement problem,
	that is, to avoid collapse in microscopic systems and produce collapse in macroscopic ones.
	
	Recent experimental work has delineating the parameter space of these models.
	Constraints on the parameter space of the CSL model have been made
	by observations on ultracold microcantilevers \citep{vinante2017, vinante2020};
	from observation of gravitational waves \citep{carlesso2016};
	matter-wave interference \citep{arndt2014};
	and neutron stars \citep{tillroy2019}.
	Spontaneous collapse models need to accommodate quantum teleportation over large distances---the record set in 2017 is 1400 km \citep{ren2017}---and
	entanglement of `macroscopic' objects up to  $10^{12}$ atoms \citep{kotler2021, lepinay2021}.
	%Quantum teleportation over large distances--the record as of 2020 is 1400 km \citep{ren2017}---and
	%`macroscopic' entanglement, where objects up to $10^{12}$ atoms have been entangled \citep{kotler2021, lepinay2021},
	%also place constraints on physical collapse models.
	In addition, spontaneous collapse gives rise to energy non-conservation \citep{pearle2000}, which is in principle observable
	as excess thermal noise \citep{adler2018, bahrami2018, mishra2018}.
	A good summary of the parameter constraints is provided by \citet{toros2018}.
	Although these constraints are directed at particular models of wavefunction collapse,
	they also provide limits on the parameters of CCQM.
	
	Note that according to this model,
	non-interacting particles will not form combined wavefunctions except in the presence nearby of particles that do interact.
	This means that one place to look for observational constraints will be in Bose-Einstein (or Fermi-Dirac) condensates,
	where there are large numbers of weakly interacting particles.
	
	Photons do not undergo collapse directly in the more recent GRW/CSL models
	but only indirectly through entanglement with matter that undergoes collapse.
	However, in CCQM photons can collapse so additional constraints on the parameters will be provided by observation of distant stars
	since the photons from such stars will undergo (multiple) collapse(s) on their path to Earth.
	A collapse of a photon wavefunction to a small volume causes a large spread in the distribution of its momentum at the point the collapse occurs,
	in accordance with Heisenberg’s Uncertainty Principle.
	Consequently, if the collapse volume were too small, it would lead to a measurable
	broadening of the image of the star.
	
	In CCQM, a small effect on the cosmic microwave background would also be expected
	due to the collapse of photon wavefunctions since the era of decoupling in the early
	universe. Constraints on the collapse parameters due to the broadening of star images
	and the effect on the cosmic microwave background are given in the appendices.
	%[removed to maintain anonymity].
	%\citet{leckey2023}.
	
	\section{Conclusion}
	\label{Sec:Conclusion}
	The outlines of a new modified form of quantum mechanics, Critical Complexity Quantum Mechanics, have been presented.
	The theory embraces significant departures from orthodox quantum mechanics
	introducing a mechanism for spontaneous collapse of the wavefunction.
	In orthodox quantum mechanics,
	we often write down wavefunctions representing a single or finite number of particles but these are approximations.
	The assumption is that the interaction of the system with the environment is small enough that the wavefunction of the system can be treated as separate from the environment and the rest of the Universe.
	In the model presented here, wavefunctions strictly represent finite numbers of particles,
	supporting a realist interpretation of the wavefunction.
	In our model,
	there is a critical relative volume which provides for an upper limit to the volume of wavefunctions in configuration space.
	Rules for the merging and splitting of wavefunctions are introduced that do not exist in orthodox quantum mechanics.
	The existence of a critical volume follows naturally from a fully discrete physics, where all physical quantities take only discrete values.
	%Although the motivation for the theory is a discrete theory of nature,
	%we argue the collapse theory can stand even if nature is continuous.
	
	The theory provides a solution to the measurement problem.
	%and one that does not suffer from the issues with tails that trouble other collapse models.
	The solution relies on the fact that where there are many particles that strongly interact
	there will be many particles per wavefunction.
	The relative volume grows exponentially with the number of particles represented,
	quickly reaching the critical volume,
	and hence collapsing, preventing systems of macroscopic size from entering into observable superpositions.
	%so collapse will happen more readily for such systems.
	
	Some advantages of the new model of collapse are that it does not suffer from the tails problem or the problems with symmetrization that afflict other collapse models.
	We suggest that CCQM provides an intuitively satisfactory image of quantum reality,
	compatible with Bells ‘density of stuff’ interpretation of the wavefunction.
	
	A fuller development of the theory,
	including the discretized Srchr\"{o}dinger dynamics
	and the rules for merging, splitting and collapsing of wavefunctions,
	is required in order to provide detailed empirical predictions.
	
	A central aim of the scientific endeavor is to provide an accurate representation of nature.
	Given that goal, it is well worth the price of introducing new laws governing the merging, splitting and collapsing of wavefunctions in order to obtain a satisfactory realist image of the quantum realm.
	These new laws generate testable predictions;
	therefore, they are worth further articulation and testing of those new predictions.
	%(see \citet[Appendices]{leckey2023})

\appendix

\section*{Appendix A --- Light from a distant star}
\label{App:Star}
In order to put some numerical constraints on the theory,
consider two situations where a one particle wavefunction may arise.
First, suppose that a photon emitted from a star
and travelling in interstellar space can be represented by a single-particle wavefunction.
According to CCQM, this photon will undergo a number of spontaneous collapses en route
to our detectors.
On each collapse to a volume of width $\Delta x$,
the photon will get a small `kick' of transverse momentum
due to Heisenberg's Uncertainty Principle:
\begin{equation}
	\Delta p = \frac{h}{4 \pi \Delta x}.
\end{equation}
Following \citet[pp.\ 158--159]{messiah1967} we assume that after emission
from an excited atom
the photon wavefunction occupies a spherical shell of constant thickness $t$.\footnote{
	\citet{messiah1967} assumes that the emitting atom is known to be in an excited state at precisely time $t=0$.
	Photon emission is a Poisson process
	and so the probability of emission as a function of time can be computed
	and from this the form of the wavefunction can be derived.
	A more complete treatment with an unknown time for the excitation of the emitting atom
	will result in the photon wavefunction having a Gaussian profile.}
Messiah gives an approximate thickness of $c \tau$, where $\tau$ is the lifetime of the atomic excited state
that emitted the photon.
We could also consider $t$ to be the coherence length of the photon.
For the moment, the exact value of $t$ is not important;
we consider it further in Appendix B.
We will assume that each collapse reduces the volume of the wavefunction by a factor of 2.
The first collapse changes the volume occupied by the wavefunction from a spherical shell to a hemispherical one.
Subsequent collapses reduce the solid angle subtended by the shell by a factor of 2.
After a number of collapses, the volume occupied by the wavefunction will be approximately
a circular disk of thickness $t$.
The first collapse occurs when the photon reaches a distance $r_c$ from the star,
when the wavefunction occupies a volume equal to the critical volume:
\begin{align}
	\label{Eq:RadiusCritical}
	V_c &= 4 \pi \, r_c^2 \, t \\ \nonumber
	{\rm or} \;\; r_c &= \sqrt{\frac{V_c}{4 \pi \, t}} \, .
\end{align}
The next collapse occurs when the volume occupied by the wavefunction again reaches $V_c$, that is, when the shell reaches a distance from the star of $\sqrt{2} \, r_c$ and so on.
If the distance from the star to the observer is $d$ and the number of wavefunction collapses
in the photon's journey to our instruments is $n$, then
\begin{align}
	\label{Eq:RadiusCriticalN}
	d &= (\sqrt{2})^{n-1} \, r_c \, , \\ \nonumber
	\Rightarrow r_c &= d \, 2^{(1-n)/2} \, .
\end{align}

The solid angle subtended by a spherical surface at its centre is given by
\begin{equation}
	\Omega = \int \int \sin \theta' d\theta' d\phi.
\end{equation}
For a cone with apex angle $2 \theta$ this becomes
\begin{align}
	\Omega_{\theta} &= \int_0^{2 \pi} \int_0^{\theta} \sin \theta' d\theta' d\phi \\ \nonumber
	&= 2 \pi (1 - \cos \theta) \\ \nonumber
	&= 4 \pi \sin^2 \left(\frac{\theta}{2} \right).
\end{align}

For large distances from the star, the half-angle subtended by the photon wavefunction shell is small, so we can make the approximation
$\cos(\theta/2) \cong 1$ implying $\sin \theta \cong 2 \sin(\theta/2)$.
In addition, note that each time the solid angle halves, $\sin(\theta/2)$ goes down by a factor of $\sqrt{2}$ (starting from $\sin(\theta/2) = 1/2$) while the radius of the shell goes up by a factor $\sqrt{2}$.
The result is that the transverse spread of the photon wavefunction, $\Delta x = r \sin \theta$,
becomes constant.
That is, for the $n$th collapse
\begin{align}
	\label{Eq:Deltap}
	\Delta x = (\sqrt{2})^{n-1} \, r_c \sin \theta
	&\cong (\sqrt{2})^{n-1} \left( 2 \left(\frac{1}{\sqrt{2}} \right)^{n+1} \right) r_c = r_c \\ \nonumber
	\Rightarrow \Delta p &\cong \frac{h}{4 \pi r_c}.
\end{align}
Hence each collapse,
apart from the first one which changes the volume occupied by the wavefunction from a spherical shell to a hemispherical one,
confines the photon in the transverse direction to an approximately
circular shell of radius $r_c$ and thickness $t$,
and each collapse contributes the same amount to the transverse momentum of the photon
through Heisenberg's Uncertainty Principle.
For the moment, since we are only after an approximate figure,
we shall ignore the fact that the earlier collapses contribute a slightly lower momentum kick
since the wavefunction shape is more curved
and so $\Delta x$ is smaller.
A more exact treatment that takes into account the different $\Delta p$ in the early collapses produces no significant divergence from the following results.

The angular deviation of the photon from a straight line path can be written as
\begin{equation}
	\Delta \phi = \frac{\Delta p_{total}}{p} = \frac{\lambda \Delta p_{total}}{h}.
\end{equation}
Since there are $n-1$ collapses that contribute to an increase in the transverse momentum of the photon,
using (\ref{Eq:Deltap}) gives
\begin{equation}
	\Delta \phi \cong \frac{(n-1) \lambda}{4 \pi r_c} \, ,
\end{equation}
and substituting for $r_c$ from (\ref{Eq:RadiusCriticalN}):
\begin{align}
	\label{Eq:NmbrCollapses}
	\Delta \phi &\cong \frac{(n-1) \, \lambda 2^{\frac{n-5}{2}}}{\pi d} \\ \nonumber
	\Rightarrow \log (n-1) + \frac{n-5}{2} \log 2 &= \log \frac{\pi d \Delta \phi}{\lambda} \, .
\end{align}
For starlight, the best resolution so far obtained is 2 milli-arc-seconds or $9.7 \times 10^{-9}$ radians
from the Very Large Telescope Interferometer for the red giant star  $\pi^1$ Gruis
which is at a distance of 530 light-years \citep{paladini2018}.
Since this is a red giant, assume $\lambda = 6 \times 10^{-7} {\rm m}$ making the right hand side of
(\ref{Eq:NmbrCollapses}) equal to $\log \, (2.5 \times 10^{17})$.
This resolution sets an upper limit for the number of collapses, $n$.
If there are more collapses, the additional transverse momentum the photon obtains
will cause an excessive deviation of the starlight and consequent blurring of the details observed.
A numerical solution to (\ref{Eq:NmbrCollapses}) gives
\begin{equation}
	%\log n + \frac{n-5}{2} \log 2 &\le& \log (2.5 \times 10^{17}) \\ \nonumber
	n \le 107
\end{equation}
Using the maximum value of $n$ in (\ref{Eq:RadiusCriticalN}) gives a minimum value for $r_c$:
\begin{equation}
	\label{Eq:StarLight_rc}
	r_c \ge \frac{d}{2^{53}} = 544 \, {\rm m}.
\end{equation}
The value of $r_c$ is both the distance from the star to the point of the first wavefunction collapse
and the radius of the (approximate) circular disk in which the photon wavefunction is confined.
The constraint on the radius can be converted to a constraint on the critical volume, $V_c$,
once a value for the thickness of the photon wavefunction has been decided
(see Equation (\ref{Eq:LimitVcStar}).

\section*{Appendix B --- Cosmic microwave background}
\label{App:CMB}
Apart from small inhomogeneities of around one part in 100,000 \citep{smoot1992},
the cosmic microwave background (CMB) is an almost perfect blackbody spectrum
with a temperature of $T = 2.725 {\rm K}$ \citep{fixsen2011}.
The frequency spectrum follows Planck's Law,
which can be written in terms of a volume energy density per unit frequency interval
\citep[p.\ 244]{griffith2005}:
\begin{equation}
	U_E(\nu) = \frac{8 \pi h \nu^{3}}{c^3} \, \frac{1}{\exp(h \nu/k_B T) - 1} \, .
\end{equation}
Dividing by the energy per photon, $h \nu$
gives the photon number density per unit frequency interval
\begin{equation}
	\label{Eq:NmbrDensity}
	N(\nu) =  \frac{8 \pi \nu^{2}}{c^3} \, \frac{1}{\exp(u) - 1} \, ,
\end{equation}
where $u = h \nu / k_B T$ is a dimensionless variable with $u \cong 2.82$ representing
the peak of the volume energy density per unit frequency curve.
For the exponential part of (\ref{Eq:NmbrDensity}),
consider, for small $\epsilon > 0$:
\begin{align}
	\exp(x(1+ \epsilon)) &= 1 + x(1 +\epsilon) + \frac{x^2(1 + \epsilon)^2}{2!} + \dots \\ \nonumber
	&= \left(1 + x + \frac{x^2}{2!} + \ldots \right) + \epsilon \, x \left(1 + x + \frac{x^2}{2!} + \ldots \right) + O(\epsilon^2) \\ \nonumber
	&\approx e^x (1 + \epsilon\,  x) \, ,
\end{align}
where the last line is obtained by dropping terms of $O(\epsilon^2)$.
Hence,
\begin{align}
	\frac{1}{\exp(x(1 + \epsilon))-1} &\approx \frac{1}{(e^x - 1) + \epsilon \, x \, e^x} \\ \nonumber
	&\approx \frac{1}{e^x - 1} \left(1 - \frac{\epsilon \, x \, e^x}{e^x - 1} \right) \, ,
\end{align}
where we have used $(1 + d)^{-1} \approx 1 - d$ for small $d$.
Hence, following \citet{pearle2018} Equation (3.34),
we can write the photon number density (\ref{Eq:NmbrDensity}) for a slightly altered temperature
$T' = T (1 - \delta)$, where $\delta << 1$, as:
\begin{align}
	\label{Eq:NmbrDensity2}
	\tilde{N}(\nu) &\approx \frac{8 \pi \nu^{2}}{c^3} \, \frac{1}{\exp(u(1+ \delta)) - 1} \\ \nonumber
	&\approx N(\nu) \left[1 - u \, \delta \frac{\exp(u)}{\exp(u) - 1} \right] \, .
\end{align}
The collapse of the photon wavefunction will introduce a slight increase in photon energy,
and hence frequency, due to Heisenberg's Uncertainty relation, as detailed in the previous section:
\begin{align}
	\label{Eq:DeltaNu}
	%\Delta p &= \frac{h}{4 \pi r_c} = \frac{h}{2 \pi} \sqrt{\frac{\pi t}{V_c}} \, , \\ \nonumber
	\Delta p &= \frac{h}{4 \pi r_c} \, , \\ \nonumber
	%\Rightarrow \Delta \nu &= \frac{c \Delta p}{h} = \frac{c}{2}  \sqrt{\frac{t}{\pi V_c}} \, .
	\Rightarrow \Delta \nu &= \frac{c \Delta p}{h} = \frac{c}{4 \pi r_c} \, .
\end{align}

Replacing $\nu$ by $\nu + \Delta \nu$ and dropping terms of $O(\Delta \nu^{2})$
the photon number density becomes
\begin{align}
	\tilde{N}(\nu) &\approx \frac{8 \pi (\nu^{2} + 2 \nu \Delta \nu) }{c^3} \, \frac{1}{\exp(u(1+ \frac{\Delta \nu}{\nu})) - 1} \\ \nonumber
	& \approx N(\nu) \left( 1 + \frac{2 \Delta \nu}{\nu} \right) \left[ 1 - u \frac{\Delta \nu}{\nu} \frac{\exp(u)}{\exp(u) - 1} \right] \\ \nonumber
	&\approx N(\nu) \left[1 - \frac{\Delta \nu}{\nu} \left(1 - 2 \frac{\exp(u)-1}{u \, \exp(u)} \right) u \frac{\exp(u)}{\exp(u) - 1} \right] \, .
\end{align}
By comparison with (\ref{Eq:NmbrDensity2}), the term
\begin{equation}
	\frac{\Delta \nu}{\nu} \left(1 - 2 \frac{\exp(u)-1}{u \, \exp(u)} \right) \, ,
\end{equation}
acts like a fluctuation (drop) in temperature.
Substituting for $\Delta \nu$ from (\ref{Eq:DeltaNu}) 
and writing the result in terms of the wavelength,
we can write the temperature fluctuation as
\begin{equation}
	\label{Eq:Delta}
	\delta = \frac{\lambda}{4 \pi r_c}  \left(1 - 2 \frac{\exp(u)-1}{u \, \exp(u)} \right) \, .
\end{equation}

Now consider two models:
(A) where the discrete cell size, $a$ is fixed,
or (B) where it is proportional to the wavelength of the particle, that is, $a = \tilde{a} \lambda$,
where $\tilde{a}$ is a constant of CCQM.
The CMB spectrum is known to some precision for approximately a factor of 10
on either side of the peak of the volume energy density per unit frequency \citep{mather1999},
so fix the relative temperature fluctuation, $\delta$
to be the maximum tolerable at a wavelength ten times that of the peak
(that is, at $\lambda \cong 19 \, {\rm mm}, {\rm or} \; u \cong 0.282$):
\begin{align}
	\label{Eq:CMB_rc}
	\delta &< 10^{-5} \\ \nonumber
	%\Rightarrow V_c &> \lambda^{3} \frac{\tilde{t}}{4 \pi} 10^{10} \left(1 - 2 \frac{\exp(u)-1}{u \, \exp(u)} %\right)^2 = 9.0 \times 10^{10} \, {\rm m}^{3}
	\Rightarrow r_c &> \frac{\lambda}{4 \pi \, \delta} \left(1 - 2 \frac{\exp(u)-1}{u \, \exp(u)} \right)
	= 112 {\rm m} \, .
\end{align}
This is smaller than the value obtained in Appendix A.
However, due to the longer photon wavelength and consequent greater value of $t$,
it leads to a stronger constraint on the critical volume.
As a first approximation,
we will take the thickness of the wavefunction to be proportional to the wavelength
of the photon: $t = \tilde{t} \lambda$.
To fix a value of $\tilde{t}$,
consider a typical visible light excited state with a lifetime of $\tau = 10^{-8}$ sec.
The full width at half maximum of the Gaussian wavepacket will be approximately $c \tau = 3 \, {\rm m}$.
The width over which the photon wavefunction falls to the threshold magnitude
permitted in a discrete physics will be several times this value.
We will estimate the width of a typical visible light photon wavefunction ($\lambda = 500 \, {\rm nm}$)
to be five times the full width at half maximum, that is, $t = 15 \, {\rm m}$,
giving a value of $\tilde{t} = 3 \times 10^7$.
The limits on the radius of the photon wavefunction shell
obtained in (\ref{Eq:StarLight_rc}) and (\ref{Eq:CMB_rc})
can now be converted to limits on the critical volume:
\begin{subequations}
	\label{Eq:LimitVc}
	\begin{align}
		\label{Eq:LimitVcStar}
		V_c &> 4 \pi \, r_c^2 \, \tilde{t}\, \lambda = 6.7 \times 10^{7} {\rm m}^{3} &({\rm starlight})\, , \\
		\label{Eq:LimitVcCMB}
		V_c &> 9.0 \times 10^{10} {\rm m}^{3} &({\rm CMB}) \, ,
	\end{align}
\end{subequations}
for the deviation of starlight and the CMB, respectively.
The constraint on $V_c$ from the observation of the cosmic microwave background
is three orders of magnitude stronger
than that obtained from the deviation of starlight from a distant star.

The constraint on the critical volume is comparable with the volume occupied by a two-photon entangled pair
used in large scale quantum teleportation experiments.
In the experiment of \citet{ren2017},
an entangled pair of photons is generated on a satellite
and distributed to two ground stations.
With a beam divergence of $10 \, \mu {\rm rad}$
and a maximum separation of the ground stations of $2400 \, {\rm km}$,
the volume occupied by the the two-photon wavefunction is approximately $1.4 \times 10^8 \, {\rm m}^3$.

Now, consider model (B).
Since the wavefunction is of a single particle we can relate the critical volume in position space, $V_c$
to the critical relative volume in configuration space, $v_c$ by
\begin{equation}
	V_c = a^3 v_c = \lambda^{3} \tilde{a}^{3} v_c \, .
\end{equation}
Using ({\ref{Eq:RadiusCritical}), we can write the temperature fluctuation in this model as
	\begin{align}
		\delta_{var} &=  \frac{\lambda}{4 \pi} \sqrt{\frac{\tilde{t} \, \lambda}{2 \pi \, \lambda^{3} \, \tilde{a}^3 v_c}}  \left(1 - 2 \frac{\exp(u)-1}{u \, \exp(u)} \right) \\ \nonumber
		&= k \left(1 - 2 \frac{\exp(u)-1}{u \, \exp(u)} \right) \, ,
	\end{align}
	where
	\begin{equation}
		\label{Eq:k}
		k = \sqrt{ \frac{\tilde{t}}{4 \, \pi \, \tilde{a}^{3} v_c} }
	\end{equation}
	is a constant dependent only on the parameters of CCQM.
	Using the value of $V_c$ from (\ref{Eq:LimitVcCMB}) and a wavelength of $\lambda = 19 {\rm mm}$,
	\begin{equation}
		\tilde{a}^{3} v_c = \frac{V_c}{\lambda^{3}} = 1.3 \times 10^{16} \, ,
	\end{equation}
	and hence $k = 1.3 \times 10^{-5}$.
	The frequency of the CMB radiation and its temperature
	both increase in proportion to redshift,
	so the parameter $u$, which depends on their ratio, is independent of the cosmological epoch.
	Hence, the temperature fluctuation induced by photon wavefunction collapse in this model
	is also independent of epoch.
	
	\begin{figure}
		\begin{center}
			\includegraphics[width=12cm]{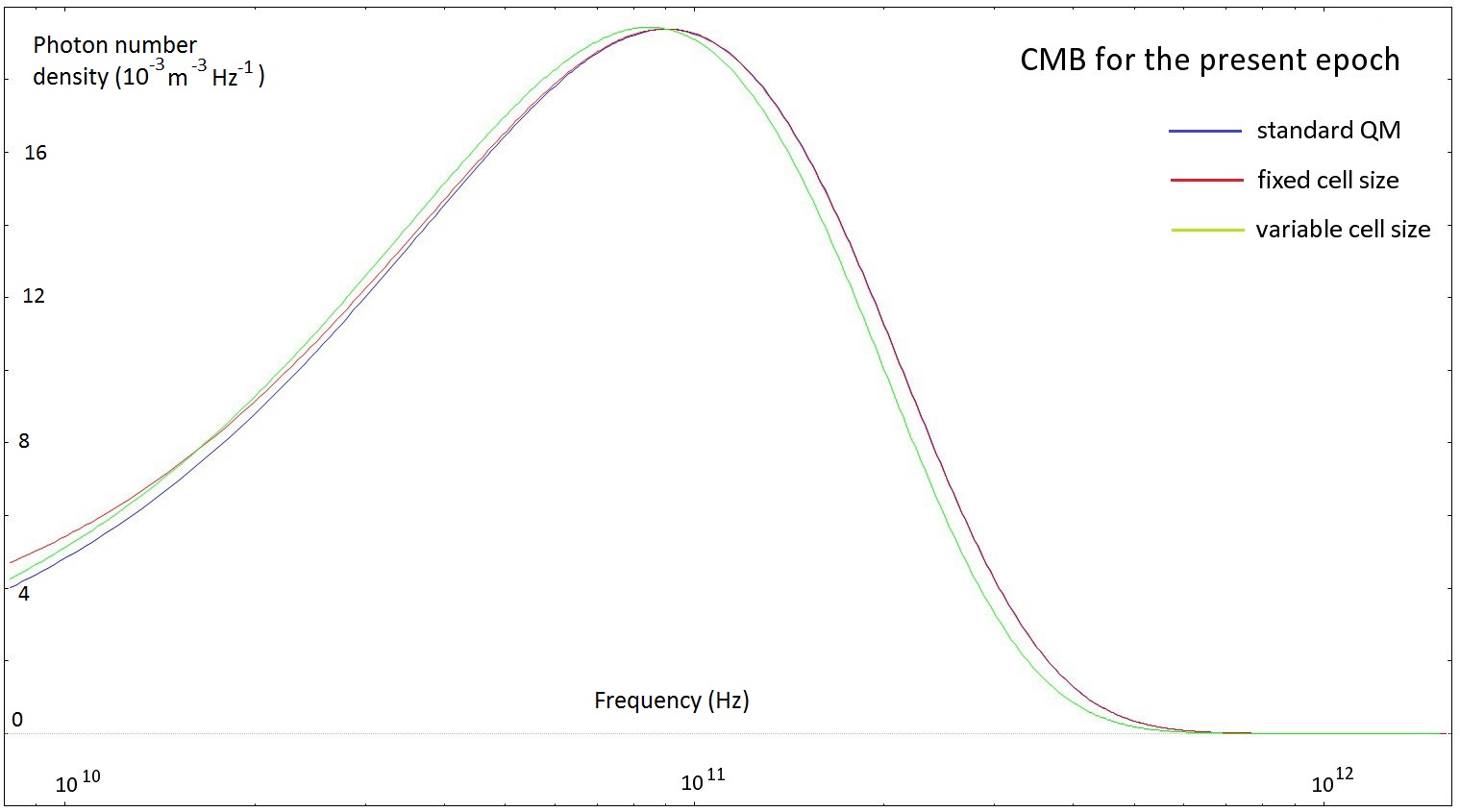}
		\end{center}
		\caption{\small Photon number density per unit frequency versus frequency
			for the Cosmic Microwave Background spectrum at the present epoch for standard quantum mechanics (blue)
			and CCQM with a fixed cell size (red) or variable cell size (green).
			The differences between standard quantum mechanics and CCQM have been magnified by a factor of 5000
			in order to be visible in the figure.
			Even at this magnification, above the peak in the photon number density ($10^{11} {\rm Hz}$),
			the difference between standard quantum mechanics
			and CCQM with a fixed cell size is not visible.}
		\label{Fig:NDensityPresent}
	\end{figure}
	
	The divergence of the spectrum from the Planck black body law
	differs between the variable and fixed cell size models
	as indicated in Figure \ref{Fig:NDensityPresent}.
	With fixed cell size, there is a divergence at low frequencies (short wavelengths),
	while the spectrum for the variable cell size case shows a systematic bias,
	the CMB temperature being higher at frequencies below that of the peak energy density
	and lower at frequencies above that of the peak.
	The spectrum for the fixed cell size case and that for standard quantum mechanics
	become increasing close as the CMB temperature rises.
	The differences between the predictions of these models becomes unimportant at earlier epochs
	where we expect most collapses to have occurred.
	However, for the variable cell sized model,
	the divergence of the CMB profile to that predicted by standard quantum mechanics is independent of epoch.
	
	The collapse of the photon wavefunction and the consequent increase in photon energy
	moves photons from one frequency interval to a higher interval.
	In the variable cell size case, below the peak number density ($u \cong 2.82$),
	there is an increase in the photon number density,
	while the situation is reversed above the peak.
	In the fixed cell size case,
	the change in the photon number density increases with decreasing frequency,
	leading to an increasing discrepancy with the Planck black body spectrum of standard quantum mechanics in the low frequency range.
	Above the peak photon number density, the curves are indistinguishable.
	
	The above analysis applies to a single collapse.
	However, since the decoupling epoch around 370,000 years after the Big Bang there would have been many collapses,
	so it is necessary to sum the change in the photon number spectrum over all such collapses.
	Using (\ref{Eq:RadiusCritical}) to substitute for $r_c$ in (\ref{Eq:CMB_rc}) gives
	\begin{equation}
		\label{Eq:Delta_z}
		\delta = \frac{\lambda^{3/2}}{4 \, \pi} \sqrt{\frac{4 \pi \, \tilde{t}}{V_c}} \left(1 - 2 \frac{\exp(u)-1}{u \exp(u)} \right)
	\end{equation}
	For earlier epochs, the wavelengths of the CMB are reduced by a factor of $z+1$,
	where a redshift of $z=0$ represents the current epoch.
	Thus, for the fixed cell size case (\ref{Eq:Delta_z}) can be written as a function of $z$ as
	\begin{equation}
		\label{Eq:Delta_z2}
		\delta(z) = -3.0 \, \sqrt{\frac{1}{V_c (z+1)^{3}}} \,,
	\end{equation}
	for a frequency of one-tenth that of the peak energy density per unit frequency ($u \cong 0.282$).
	Decoupling occurred at a redshift of approximately $z = 1100$.
	The total deviation from the Planck black body spectrum will be the sum of $\delta(z)$
	over all the redshifts $z$ at which a collapse occurred.
	Since $\delta(z)$ scales as $z^{-1.5}$ the contributions from earlier times
	become increasingly small and are negligible close to the decoupling era.
	Limiting $\sum \delta(z)$ to $10^{-5}$, the constraint on the critical volume becomes
	\begin{equation}
		\label{Eq:VcRedshift}
		V_c > 9.0 \times 10^{10} \sum\limits_{z} (z+1)^{-3} \, . 
	\end{equation}
	Most of the collapses will occur in the early universe,
	and these will make only a small contribution to the sum in Eq.\ (\ref{Eq:VcRedshift}),
	so this constraint will be weaker than the one obtained in the variable cell size model,
	where the deviation from the Planck spectrum generated by a collapse is independent of the CMB temperature.
	Although the fixed cell sized model gives rise to smaller deviations
	from the Planck spectrum,
	the model presents other problems.
	In particular, the cell size must be much smaller than the shortest wavelength particle
	that we have observed, in order for noticeable effects not to be apparent
	in other experiments.
	%Since $V_c = a^{3} v_c$, the smallness of $a$ implies a very large value for the relative critical volume $v_c$
	%for a given value of the critical volume, $V_c$.
	
	Although the calculations presented above are only approximate,
	they present patterns of deviation from the Planck spectrum that could be observed in future experiments.

\end{document}